# On the short-term influence of oil price changes on stock markets in GCC countries: linear and nonlinear analyses

**Mohamed El Hedi AROURI**

(LEO-Université d'Orléans & EDHEC, mohamed.arouri@univ-orleans.fr)

**Julien FOUQUAU**

(ESC Rouen &  LEO, julien.fouquau@univ-orleans.fr)

**Abstract**

This paper examines the short-run relationships between oil prices and GCC stock markets. Since GCC countries are major world energy market players, their stock markets may be susceptible to oil price shocks. To account for the fact that stock markets may respond nonlinearly to oil price shocks, we have examined both linear and nonlinear relationships. Our findings show that there are significant links between the two variables in Qatar, Oman, and UAE. Thus, stock markets in these countries react positively to oil price increases. For Bahrain, Kuwait, and Saudi Arabia we found that oil price changes do not affect stock market returns.

# 1. Introduction

Over the last three decades, oil prices have changed with sequences of very large increases and decreases. As illustration of this, the oil price increased by 76% between March 2007 and June 2008 and then decreased by 48% between July 2008 and October 2008. Thus, it sounded natural to have a look at the impact of oil prices on the macroeconomic variables. Most studies have established significant effects of oil price changes on economic activity in several developed and emerging countries [Cunado and Perez de Garcia (2005), Balaz and Londarev (2006), Gronwald (2008), Cologni and Manera (2008), and Kilian (2008)].

In contrast with the volume of work done investigating the link between oil price shocks and economic activity, there has been relatively little work done on the relationships between oil price variations and stock markets, as underlined by Basher and Sadorsky (2006). Furthermore, most of these works have focused on few industrial countries, mainly the US, Canada, Europe, and Japan [Jones and Kaul (1996), Huang *et al.* (1996) and Sadorsky (1999)]. The results of these studies are inconclusive. Recently, some papers have focused on major European, Asian and Latin American emerging markets. They have shown a significant short-term link between oil price changes and these emerging stock markets. For instance, using a VAR model, Papapetrou (2001) shows a significant relationship between oil price changes and stock markets in Greece. Basher and Sadorsky (2006) use an international multifactor model and reach the same conclusion for other emerging stock markets.

However, less attention has been paid to smaller emerging markets, especially in the Gulf Cooperating Council (GCC) countries where share dealing is a relatively recent phenomenon. Conversely, studying the link between oil prices and stock markets in GCC countries is interesting for several reasons. First, GCC countries are major suppliers of oil in world energy markets, their stock markets are likely to be susceptible to change in oil prices. Second, the GCC markets differ from those of developed and from other emerging countries in that they are largely segmented from the international markets and are overly sensitive to regional political events. Finally, GCC markets represent very promising areas for regional and world portfolio diversification. Thus, understanding the influence of oil price shocks on GCC stock market returns is important for investors to make necessary investment decisions and for policy-makers to regulate stock markets more effectively.

As far as we know, only three recent papers have attempted to investigate the relationships between oil and stock markets in GCC countries. Using VAR models and cointegration tests, Hammoudeh and Eleisa (2004) show that there is a bidirectional relationship between Saudi stock returns and oil price changes. The findings also suggest that the other GCC markets are not directly linked to oil prices and are less dependent on oil exports and are more influenced by domestic factors. Bashar (2006) uses VAR analysis to study the effect of oil price changes on GCC stock markets and shows that only the Saudi and Omani markets have predictive power of oil price increase. Finally, Hammoudeh and Choi (2006) examine the long-term relationship of the GCC stock markets in the presence of the US oil market, the S&P 500 index and the US Treasury bill rate. They find that the T-bill rate has a direct impact on these markets, while oil prices and the S&P 500 have indirect effects.

As we can see, the results of the few available works on GCC countries are too heterogeneous. These results are puzzling because the GCC countries are heavily reliant on oil export (and thus sensitive to changes in oil prices) and have similar economic structures. The aim of this article is to add to the current literature on the subject by examining the short-term interaction between oil price changes and stock returns in GCC countries. To do so, we have first used a linear specification. In view of the results we obtained, we show that this link is not systematic for all the GCC countries. In fact, oil prices affect significantly stock markets only in Qatar, UAE, and Oman.

However, some recent papers have shown that the link between oil and economic activity is not entirely linear and that negative oil price shocks (price increases) tend to have larger impact on growth than positive shocks do [Hamilton (2003), Zhang (2008), Lardic and Mignon (2006, 2008) and Cologni and Manera (2009)]. Thus, we should expect that oil prices equally affect stock markets in a nonlinear fashion, especially in GCC countries. In fact, for these net oil-exporting countries oil price fluctuations should increase corporate output and earnings, but also domestic price levels. Therefore, unlike net oil-importing countries where the expected link between oil prices and stock markets is negative (oil price increases have a negative impact on future cash-flows), the transmission mechanism of oil price shocks to GCC stock market returns are ambiguous and the total impact of oil price shocks on stock returns depends on which of the positive (positive impacts of oil prices increase on GCC firms and economies) and negative (domestic and imported inflation) effects counterweigh the other.

In a second step, we have then checked if there are signs of nonlinearities in the link between oil prices and stock markets in GCC countries. For this aim, we applied a non parametric model to allow the estimation of non-linear relationships. Despite the use of this method, we reach the same results. We find that the relationships between oil prices and stock markets returns are significant (and slightly nonlinear) for Qatar, UAE and Oman. For the three other countries, there is no relationship.

The outline of this paper is as follows. The next section describes the data and briefly presents the GCC stock markets. Section 3 discusses the preliminary results we obtained on the basis of linear specifications. Section 4 presents the non-parametric model specification and reports the results we obtained from the nonlinear methods. The last section concludes.

## 2. Data and the role of oil in GCC countries

We use weekly data for the six members of the Gulf Cooperation Council (GCC): Bahrain, Kuwait, Oman, Qatar, Saudi Arabia and United Arab Emirates (UAE). Our sample covers the period from the first week of June 2005 to the third week of October 2008, which represents 177 observations for each country. Stock market indices are obtained from MSCI. For oil, we use the weekly OPEC countries spot price weighted by estimated export volume obtained from the Energy Information Administration (EIA). OPEC prices are often used as reference prices for crude oil including oil produced by GCC countries.[1] All prices are denominated in American dollars.

Table 1 displays descriptive statistics of the stock returns in GCC countries as well as the world market returns measured by MSCI world index returns. Compared with the World market, GCC stock markets have higher volatilities, but not necessarily higher returns. Kuwait has the highest returns followed by Oman. Saudi Arabia has the highest risk followed by UAE. On average, oil price changes are higher than stock market returns. Furthermore, many average returns are negative as a result of the subprime crisis. Finally, except for Kuwait, the normality hypothesis is rejected for all the series we study.

Together, GCC countries produce about 20% of all world oil, control 36% of world oil exports and possess 47% of the world oil proven. Oil exports largely determine earnings, government budget revenues and expenditures and aggregate demand. Table 2 shows some key financial indicators for the stock markets in GCC countries. The contributions of oil to GDP range from 22% in Bahrain to 44% in Saudi Arabia. Moreover, for the three largest economies in GCC countries, Saudi Arabia, UAE, and Kuwait, the stock market's liquidity indicator is positively associated with the oil importance indicator in these economies.

---

[1] Very similar results are obtained with West Texas Intermediate and Brent spot prices. Oil prices are in US dollar per barrel. Note also that GCC currencies are pegged at fixed rates to the U.S. dollar officially since 2003.

Saudi Arabia leads the region in terms of market capitalization. However, by percentage of GDP, Qatar is the leader. Stock market capitalization exceeded GDP for all counties except Oman. Kuwait has the largest number of listed companies, followed by Oman. Overall, GCC stock markets are limited by several structural and regulatory weaknesses: relatively small numbers of listed firms, large institutional holdings, low sectoral diversification, and several other deficiencies. In recent years, however, a broad range of legal, regulatory, and supervisory changes has increased market transparency.

## 3. The link between oil and stock returns: a simple linear investigation

We investigate here if the oil price changes ($Ropec_t$) explain the stock market returns ($R_t^{(i)}$) in GCC countries. A version of this relationship may be represented by the following equation[2]:

$$R_t^{(i)} = a_i + \beta_i Vmsci_t + \delta_0 Ropec_t + \varepsilon_t \qquad (1)$$

where $\varepsilon_t$ is a stochastic error term and $Vmsci_t$ represents the world market returns filtered by the oil price returns. More precisely, $Vmsci_t$ is the residuals of the OLS regression of the world market returns ($Rmsci_t$) on the oil price changes:

$$Rmsci_t = \alpha + \beta Ropec_t + V_t \qquad (2)$$

The results are reported in Table 3. We have first tested for the assumptions of no serial correlation and homogeneity of the error terms. Serial correlation of order one[3] have been found for Bahrain. To remove it, we added in the equation (1) the Bahraini stock market returns lagged of one period. Second, we have tested for the presence of heteroskedasticity. More precisely, we have applied the White's test. We found three countries with unknown form of the heteroskedasticity: Oman, Qatar and UAE. In these cases, the OLS estimator is unbiased but inefficient. Then, we use the White's estimator to estimate the asymptotic covariance matrix.

In View of these results, there exists a positive and significant relation between oil price changes and stock market returns for three countries: Oman, Qatar, and UAE. For these countries the positive effects of oil price increases offset the negative effects and oil price changes are positively reflected in stock returns. The elasticity of stock prices to oil prices in these countries is less than one, but the stock price effect of oil changes is great: a 10% increase in oil prices leads to an average appreciation of the stock markets by 2.1%, 4.1% and 3.4% respectively fro Oman, Qatar, and UAE.

For the other countries (Bahrain, Kuwait, and Saudi Arabia), the global effects of oil price changes on stock returns is insignificant. The absence of relationships between oil prices and stock markets in too oil dependent economies such as Bahrain, Kuwait and Saudi Arabia countries seems to be counterintuitive. However, as we have previously mentioned, some recent papers have shown that the link between oil and economic activity is not entirely linear and that there is some evidence of nonlinearities between the two variables [Hamilton (2003), Zhang (2008) and Lardic and Mignon (2006, 2008)]. Therefore, one possible explanation for our findings is that the traditional linear specifications are too restrictive and cannot reproduce such nonlinearities. In the rest of the paper, we investigate whether our results are not conditioned to the assumption of a linear relationship.

---

[2] We obtain similar results if we consider an alternative equation without the filtered variable $Vmsci_t$. The results oh these estimations are available upon request.
[3] We have not found serial correlation of order superior.

## 4. Further evidence from a nonlinear investigation

Unlike the previous section, we examine here the relationships between oil prices and stock markets without imposing the assumption of linearity. More particularly, we use a non parametric method: a local polynomial kernel regression of order 2 with a Gaussian kernel, which has the advantage to do not need *a priori* functional models. However, the major drawback is that there is no analytical form of the link between the two variables with this solution.

This method consists to fit the dependent variable Y (here the stock market returns) at each value x (the grid points, here the oil price changes) by choosing the set of parameters $\beta_i$ for $i = 0, 1, 2$ to minimize the weighted sum of squared residuals:

$$SSR(x) = \sum_{i=1}^{T} (Y_i - \beta_0 - \beta_1(x - X_i) - \beta_1(x - X_i)^2) K\left(\frac{(x - X_i)}{h}\right) \quad (3)$$

where T is the number of observations of stock market returns, h is the bandwidth and K is a kernel function that integrates to 1. The kernel is the function used to weight the observations in each local regression. Observations far of the estimated point x have less impact on the estimation. There are many possible kernel functions. However, this choice does not modify significantly the results with the exception of the uniform kernel. In this study, we use a Gaussian kernel:

$$K\left(\frac{(x - X_i)}{h}\right) = K(u) = \frac{1}{\sqrt{2\pi}} \exp\left(-\frac{1}{2}u^2\right)$$

with an arbitrary bandwidth[4] $h = 0.15(X_u - X_l)$ and $(X_u - X_l)$ the range of X. The choice of the bandwidth parameter corresponds to an arbitration variance / bias (For more precision see Cleveland et Devlin (1988)).

Figure 1 displays scatter plots of the oil price changes and the stock market returns. We also report local polynomial kernel regression of order 2 in the left part. By way of comparison, we report the line of the traditional linear regression in the right part.

In view of these results, the relationship between oil prices and stock returns in GCC countries is not clearly nonlinear. It appears especially that there are some edge effects. For Qatar, Oman, and UAE, the link between oil price changes and the stock market returns is positive and the highest for negative values of oil price shocks (-0.15 to -0.5). However, the relationship remains positive but lower for higher values. This link seems therefore nonlinear and switching according to the oil price change values. The results we found for the three other countries are in agreement with those obtained using parametric linear model. With the exception of edge effect, there is no relationship between oil price changes and stock markets returns in Bahrain, Kuwait, and Saudi Arabia.

## 4. Conclusion

This paper investigated the short-run relationships between oil prices and GCC stock markets. Since GCC countries are important world energy market players, their stock markets may be susceptible to oil price shocks. To account for the fact that stock markets may respond nonlinearly to oil price shocks, we have examined both linear and nonlinear linkages. Our results show that there are significant links between the two variables in Qatar, Oman, and UAE. Thus, stock markets in these countries react positively to oil price increases. For Bahrain, Kuwait, and Saudi Arabia we found that oil price changes do not affect stock market returns.

---
[4] We use the bandwidth parameter provided by *e-views*.

Our findings should be of interest to researchers, regulators, and market participants. In particular, GCC countries as OPEC policy-makers should keep an eye on the effects of oil price fluctuations on their own economies and stock markets. For investors, the significant relationship between oil prices and stock markets implies some degree of predictability in some GCC stock markets.

There are several avenues for future research. First, the relationships between oil price changes and stock market returns in GCC countries can be expected to vary from one economic sector to another. A sector analysis of this link would be informative. Second, the same approach applied in this article could be used to examine the effects of other energy products, such as natural gaz. Third, further research could examine the links of causality binding oil and stock markets in GCC countries and other oil-exporting countries.

## Table 1: Descriptive statistics

|  | OPEC | Bahrain | Kuwait | Oman | Qatar | Saudi | U.A.E. | World |
|---|---|---|---|---|---|---|---|---|
| **Mean (*100)** | 0.226 | -0.090 | 0.240 | 0.002 | 0.026 | -0.253 | -0.346 | -0.013 |
| **Std. Dev.** | 0.036 | 0.026 | 0.028 | 0.030 | 0.040 | 0.051 | 0.043 | 0.006 |
| **Skewness** | -0.803 | 0.490 | -0.186 | -0.971 | -0.563 | -1.113 | -2.059 | -4.231 |
| **Kurtosis** | 4.693 | 6.568 | 3.237 | 8.079 | 6.866 | 7.052 | 15.040 | 36.116 |
| **Jarque-Bera** | 39.93*** | 100.39*** | 1.42 | 216.87*** | 118.93*** | 156.74*** | 1187.4*** | 8567.7*** |

*, ** and *** denote rejection of the null hypothesis respectively at 10%, 5% and 1%.

## Table 2: Stock markets in GCC countries in 2007

| Market | Number of companies* | Market Capitalization ($ billion) | Market Capitalization (% GDP) * | Oil (% GDP)+ |
|---|---|---|---|---|
| **Bahrain** | 50 | 21.22 | 158 | 22 |
| **Kuwait** | 175 | 193.50 | 190 | 35 |
| **Oman** | 119 | 22.70 | 40 | 41 |
| **Qatar** | 40 | 95.50 | 222 | 42 |
| **UAE** | 99 | 240.80 | 177 | 32 |
| **S. Arabia** | 81 | 522.70 | 202 | 44 |

*Sources: Arab Monetary Fund and Emerging Markets Database. * Numbers in 2006.*

## Table 3: Linear model results

|  | Bahrain | Kuwait | Oman | Qatar | Saudi | U.A.E. |
|---|---|---|---|---|---|---|
| $\alpha$ | -0,001 | 0,002 | -0,000 | -0,001 | -0,003 | -0,004 |
|  | ( 0,002) | (0,002) | (0,002) | (0,003) | (0,004) | (0,003) |
| $\beta$ | 0,033 | -0,028 | 0,210*** | 0,408*** | 0,141 | 0,3420*** |
|  | (0,052) | (0,057) | (0,070) | (0,108) | (0,104) | (0,130) |
| $\delta$ | 0,204*** | 0,108 | 0,303** | 0,154 | 0,296** | 0,429 |
|  | (0,073) | (0,081) | (0,150) | (0,226) | (0,147) | (0,302) |
| $\overline{R}^2$ | 0,065 | 0,001 | 0,121 | 0,139 | 0,021 | 0,142 |
| **Log-ikelihood** | 400,612 | 382,847 | 379,688 | 332,779 | 277,991 | 320,872 |
| **F-statistic** | 5,070*** | 1,010 | 13,108*** | 15,218*** | 2,921** | 15,5581*** |
| **AIC** | -4,5327 | -4,316 | -4,280 | -3,747 | -3,124 | -3,612 |

*, ** and *** denote rejection of the null hypothesis respectively at 10%, 5% and 1%. Robust standard errors are reported in parentheses.

# Figure 1: Scatter plots of the oil price returns and the stock markets returns and local polynomial kernel regression of order and linear regression

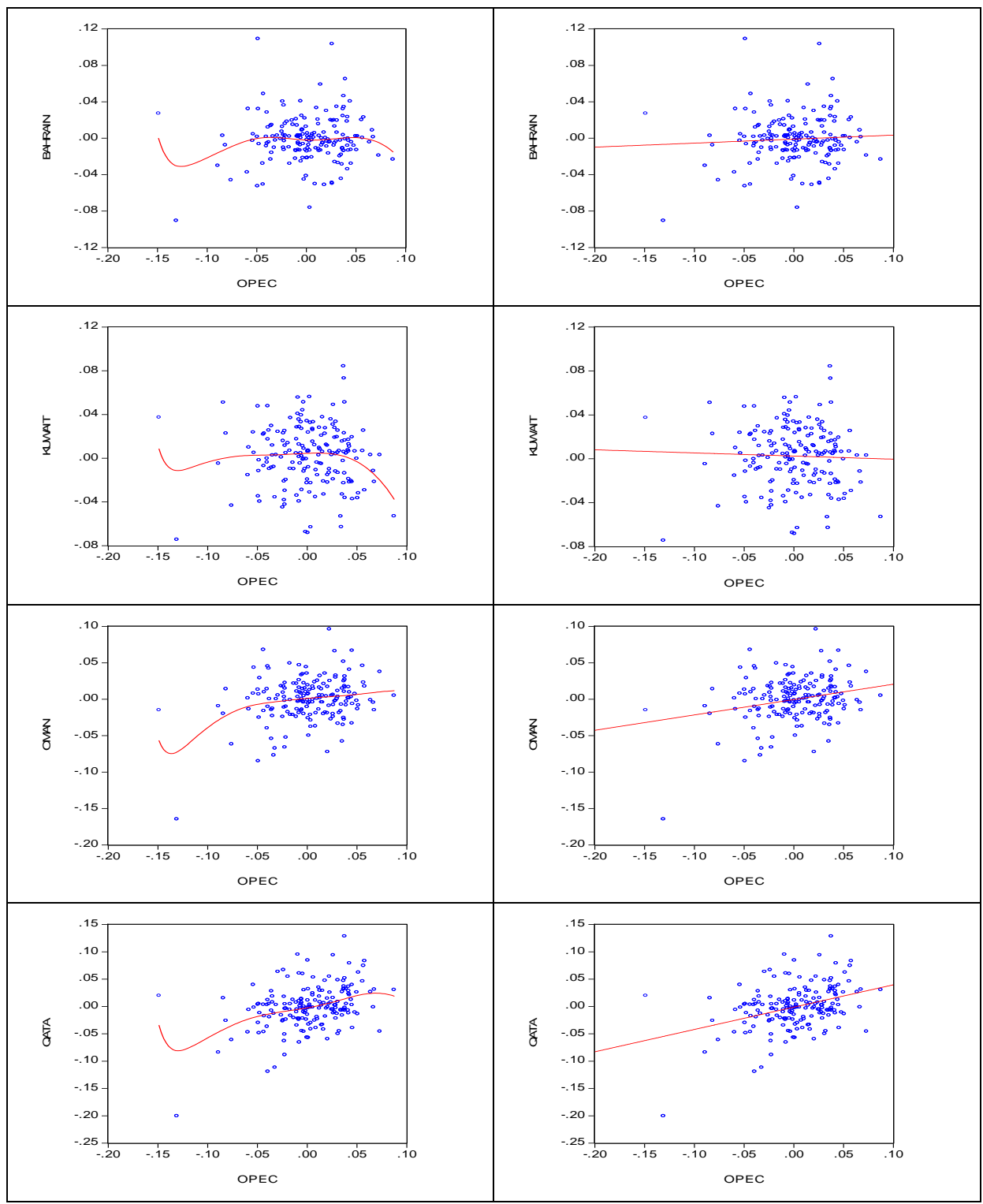

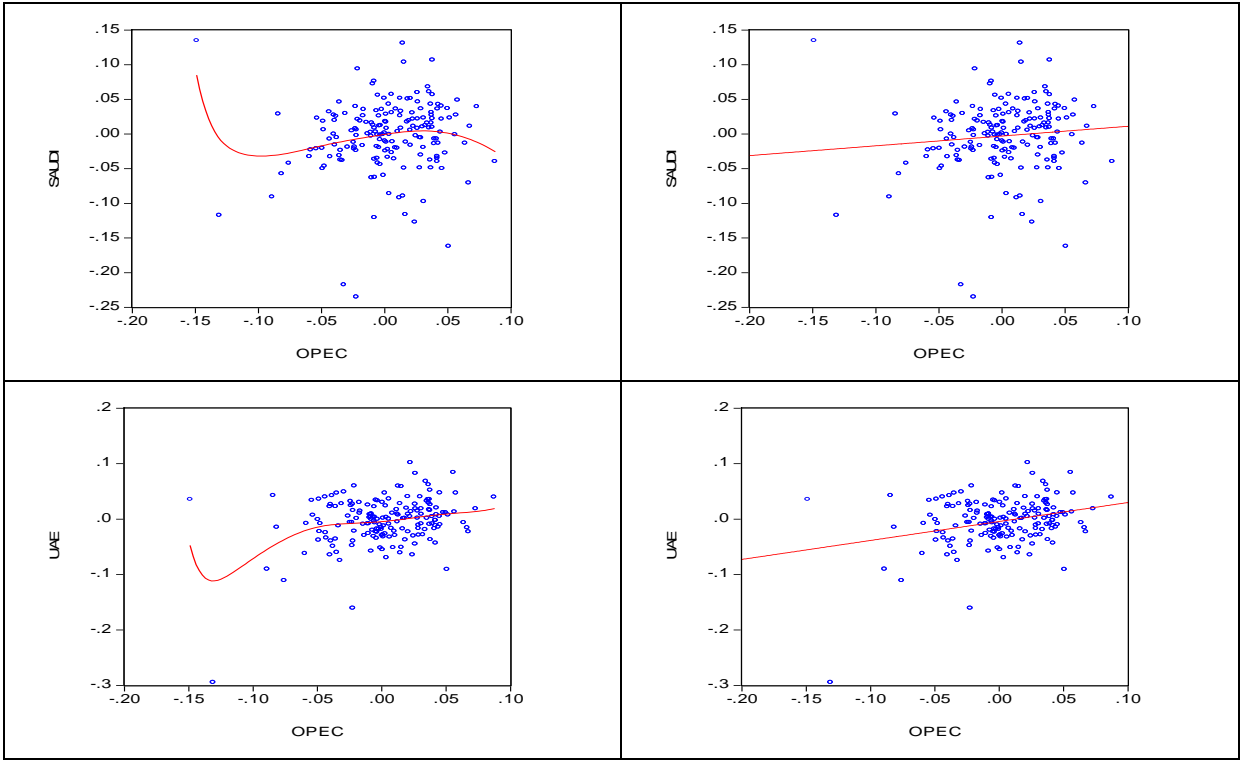